# Strong permanent magnet gradient deflector for Stern-Gerlach-type experiments on molecular beams

Jiahao Liang[1], Thomas M. Fuchs[2], Rolf Schäfer[2], and Vitaly V. Kresin[1*]

[1]Department of Physics and Astronomy, University of Southern California,
Los Angeles, California 90089-0484, USA

[2]Eduard-Zintl-Institut für Physikalische und Anorganische Chemie,
Technische Universität Darmstadt, Alarich-Weiss-Straße 8, 64287 Darmstadt, Germany

We describe the design, assembly, and testing of a magnet intended to deflect beams of paramagnetic nanoclusters, molecules, and atoms. It is energized by high-grade permanent neodymium magnets. This offers a convenient option in terms of cost, portability, and scalability of the construction, while providing field and gradient values (1.1 T, 330 T/m) which are fully comparable with commonly used electromagnet deflectors.

---

* Corresponding author. Email: *kresin@usc.edu*

# I. INTRODUCTION

The use of inhomogeneous magnetic fields for the deflection of atomic beams originated with the foundational experiment by Stern and Gerlach[1,2] and was subsequently extended to beams of molecules[3-5] and size-selected nanoclusters (see, e.g., the citations in Refs. 6,7). The magnetic deflection method provides important information about the strength, makeup, and dynamics of these particles' magnetic moments.

For both historical and practical reasons, these experimental setups have employed electromagnets. Before rare-earth permanent magnets became commercially available in the 1970s[8] this was the only practical way to generate a sufficiently strong extended deflecting field. Moreover, a reference undeflected profile can be obtained easily by switching off the current in the windings.

In 1987 Ziock and Little[9] described the construction of a compact high-gradient quadrupole focuser which was energized by permanent samarium cobalt magnets instead of wire coils. They employed it with beams of In atoms and small Bi clusters.[10] Recently, a Halbach-type deflection magnet was used in an atomic beam interferometry experiment.[11] The incentives for such designs are cost savings (e.g., no need for a stable, regulated high-current power supply) as well as compactness, simplified construction, reduced weight, and full ultrahigh vacuum compatibility (no need for windings, cooling water, feedthroughs, and/or vacuum barriers). The same motivations apply to a deflection Stern-Gerlach type magnet described in this paper. An additional advantage is portability. Indeed, the device was designed and built in Los Angeles and then easily shipped to and tested in Darmstadt.

One of the intended uses of this magnet are deflection experiments on ultra-cold superfluid helium nanodroplets with embedded (super)paramagnetic atoms, molecules, and clusters. Helpfully, in this case a reference undeflected profile can be acquired from an undoped, nonmagnetic, droplet beam without needing to switch off the field. This is analogous to recent electric deflection experiments on nanodroplets with embedded polar molecules[12,13] in which undoped nanodroplets were unaffected by passage between high-voltage deflection electrodes. The same scheme can be extended to experiments with various clusters of other types that are doped with a magnetic impurity.[14] Likewise, when certain clusters in a size sequence display negligible susceptibilities compared to their near neighbors[15,16] they may be used to track the undeflected beam even with the magnet in place. In other circumstances, the magnet can be moved out of the beam path by a precise mechanical translation stage.

The paper is organized as follows. Sec. II focuses on the mechanical construction, while Sec. III describes the selected pole shoe shape and its magnetic field configuration. Sec. IV describes an experimental test of the deflector in a Stern-Gerlach atomic beam measurement and the determination of its magnetic field gradient. Sec. V presents a summary.

### II. CONSTRUCTION

The construction is shown in Fig. 1. The field is generated by three cube-shaped nickel-plated N52 neodymium magnets of 50.8 mm edge length (SM Magnetics, Part no. C2002_N52) stacked side by side inside an aluminum square tubing (63.5 mm outer width, 4.8 mm wall thickness). The gaps between the magnets and tubing's inner wall are filled with 1 mm thick bronze strips whose purpose is to center the magnets and to assist with sliding them into the channel. The magnet insertion procedure is described later in this section.

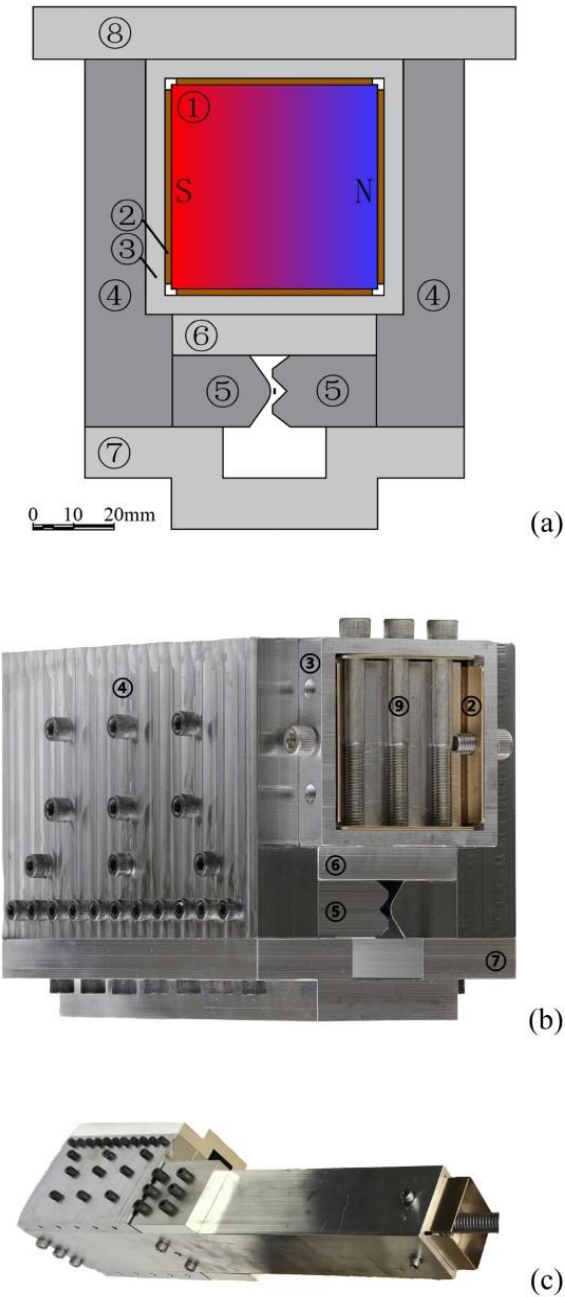

**Figure 1**. The magnetic deflection assembly's face view (a) and a photograph (b). ① Magnet stack with magnetization direction shown; ② lengthwise shim strips; ③ square tubing; ④ yoke plates; ⑤ pole shoes; ⑥ alignment spacer; ⑦ brace; ⑧ mounting plate, ⑨ magnet confinement bolts. The small black dash in the gap between the pole shoes in panel (a) marks the approximate entry position of the collimated molecular beam. Panel (c) shows the channel extension and press screw used for insertion of the permanent magnets.

The yoke plates and the pole shoes were machined out of permendur alloy plate (Hyperco-50A, acquired from Ed Fagan Inc.) and subsequently thermally annealed in vacuum (Thermal-Vac Technology). The shape of the pole shoes and of the field they produce are discussed in Sec. III. The spacer and the brace are made out of aluminum; their heights and the gap between them serve to align the pole shoes precisely relative to each other. The aluminum mounting plate on the other side adds further rigidity to the assembly and is also used to install it in the vacuum chamber. At the Darmstadt apparatus the assembly was attached to a top flange by standoffs, while at USC it is suspended from a heavy duty linear feedthrough which can raise it to allow the molecular beam to pass through the opening in the bottom brace unimpeded.

The entire framework is bolted together prior to insertion of the three permanent magnets. At this point, the challenge is to find a way to insert them one after the other into the square channel. Being extremely strong, they are guaranteed to jump out of one's hand and irretrievably stick to the yoke plates if they are simply brought up to the channel opening. At the same time, these rare-earth cubes are smooth and highly brittle, so they cannot be tightly clamped in some mechanical holder. In addition, even if they somehow made it into the channel, the aligned magnets would strongly repel each other at close distances.

This issue was solved by adding a special extension to the square tubing which allowed us to insert the magnets with minimal effort and complications. As can be seen in the photograph in Fig. 1(c), the square channel ③ extends beyond the edge of the deflector assembly by approximately 1.5 cm on each side. Using threaded holes drilled in these ends, we attached a 30 cm-long piece of identical square tubing to one of them, as shown in Fig. 1(c). The aforementioned bronze strips lining the walls of the channel were initially made long enough to reach all the way to the end of the extension. Now the cube magnets could be inserted one by one into the distant opening and pushed forward by a nonmagnetic bar. Being fully contained by the long extension channel, they easily slid along the bronze-lined walls all the way into position between the yoke plates. Since most of their magnetic flux was dissipated into the yokes, the aligned permanent magnets did not begin to repel each other until they were separated at most by a couple of centimeters, and the repulsion was notably weaker than if they had been brought together in empty space. After all three magnets were inserted, a press screw with a handle was attached to the entrance of the extension channel [see Fig. 1(c)], and used to bring them into full contact. The magnets having been fenced in by bolts [visible in Fig. 1(b)] on both ends, the extension channel was then removed and the extra length of the bronze strips cut off. An analogous arrangement can be used to push the magnets out of the assembly if needed.

Fig. 1(a) also indicates the area intended for the molecular beam. In Darmstadt, the beam is collimated to a rectangular shape of 0.2 mm width and 2.0 mm height by two slits before entering the magnet. For the USC lab we designed a short Macor "plug" with a narrow slit (0.25 mm width × 1.25 mm height) to fit tightly directly into the front gap between the pole shoes so as to correctly position and collimate the beam.

### III. MAGNETIC FIELD

The force on an atomic, molecular, or cluster beam due to an inhomogeneous magnetic field is given by $\vec{F} = \langle \mu \rangle \nabla B$, where $\langle \mu \rangle = -\partial U / \partial B$ is the time-averaged projection of the particles' effective magnetic dipole moment on the field direction ($U$ is the energy of the particle in the magnetic field; see, e.g., Refs. [4,17]). Thus, if the beam travels in the $y$ direction and the magnet is designed to have a strong gradient in the $z$ direction, the beam deflects sideways by an amount proportional to $\langle\mu\rangle\partial B/\partial z$. (The gradient ∂B/∂x broadens, or defocuses, the beam in the vertical direction.[18,19])

If $\langle\mu\rangle$ itself is proportional to the magnetic field $B$ (as is often the case with rotating molecules and with clusters exhibiting a linear magnetic susceptibility) the beam deflection becomes proportional to $B(\partial B/\partial z) \propto \partial(B^2)/\partial z$ and the magnetic pole design aims to make this quantity as uniform as possible over the beam volume. This can be accommodated by the "two-wire" or "Rabi" geometry.[20,21] On the other hand, if the magnitude of $\langle\mu\rangle$ is fixed (as is the case with atoms) or nearly saturated (as anticipated for field-oriented molecular magnets within helium nanodroplets), then uniformity of the gradient ∂B/∂z is desired. The so-called quadrupole sector geometry[5,22] is often used for this purpose, and is adopted here.

It is important to emphasize that the permanent magnet deflector construction can of course accommodate various pole configurations.

Based on extensive simulations using FEMM software,[23] we verified that the pole shape described in Ref. 18 ("4 mm gap quadrupole-like configuration") offered the optimal combination of field strength and gradient uniformity for our system. The pole profile can be seen in Fig. 1(b). Fig 2(a) shows the calculated magnetic field in the region between the pole shoes, Fig. 2(b) shows the deflecting gradient ∂B/∂z, and Fig. 2(c) shows |∂B/∂x|.

The magnetic field within the actual assembled unit was measured as follows. The tip of a Hall magnetometer probe was secured in a small aluminum slider which fit snugly between the pole shoes and could be translated along their entire length. The probe was positioned very close, in *x* and *z,* to the center of the collimated beam region. In this way, it was determined that the magnetic field value at that position between the pole shoes was constant to better than 1% along the entire 15 cm length of the magnet, and was equal to 1.1 T. This is 35% smaller than predicted by the simulation in Fig. 2(a). As described in the next section, this is consistent with the experimentally determined magnetic gradient, which is also 35% below the model value. This deviation in absolute values is attributed to a difference in materials properties between simulation and reality, for example a lower magnetization of the N52 alloy and/or incomplete annealing of the permendur pieces.

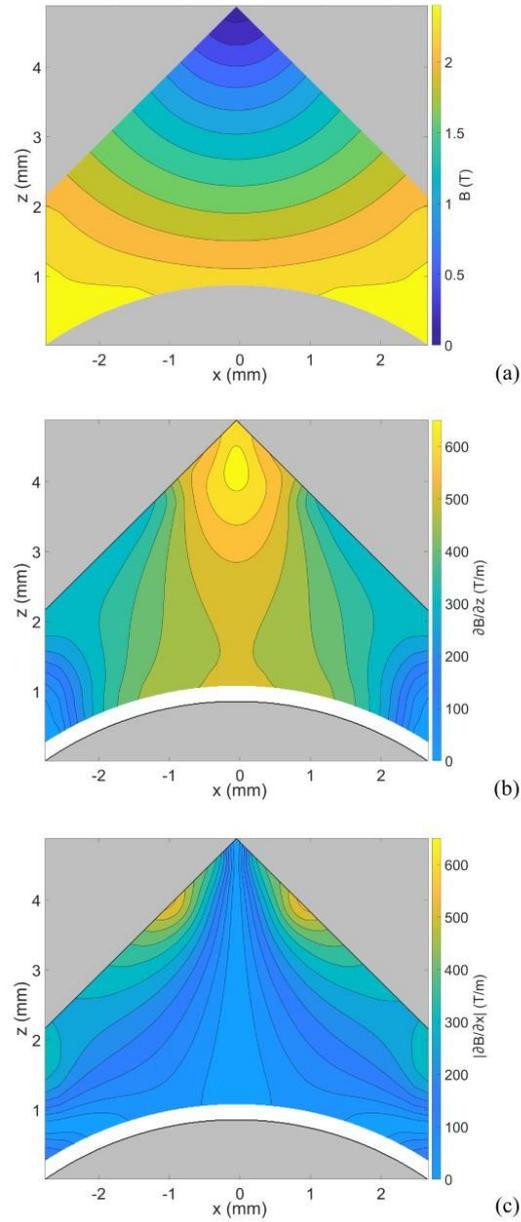

**Figure 2.** (a) The magnetic field, (b) its deflecting gradient $\partial B/\partial z$, and (c) the defocusing gradient $|\partial B/\partial x|$ between the deflector's poles. The pole gap is 4 mm and the pole radius is 4.74 mm. The contour lines in (a) are drawn at intervals of 0.2 T, and those in (b,c) at intervals of 50 T/m. The calculations were performed with the use of FEMM software and its nominal library parameters for the magnet and yoke alloys (actual material parameters vary, see text). The plots were prepared by employing 0.01 mm mesh size in the region between the pole shoes, differentiating the calculated field numerically and then smoothing the field and derivative values with a 0.2 mm window for display in this figure.

### IV. ATOMIC BEAM TEST OF THE DEFLECTION MAGNET

In order to verify the performance of the new deflector, it was installed in an apparatus where a calibrated electromagnet has been used to study the magnetic properties of atoms and clusters.[24] A Stern-Gerlach experiment on Bi atoms was carried out and compared with the electromagnet data. A sketch of the experimental arrangement is shown in Fig. 3.[25,26]

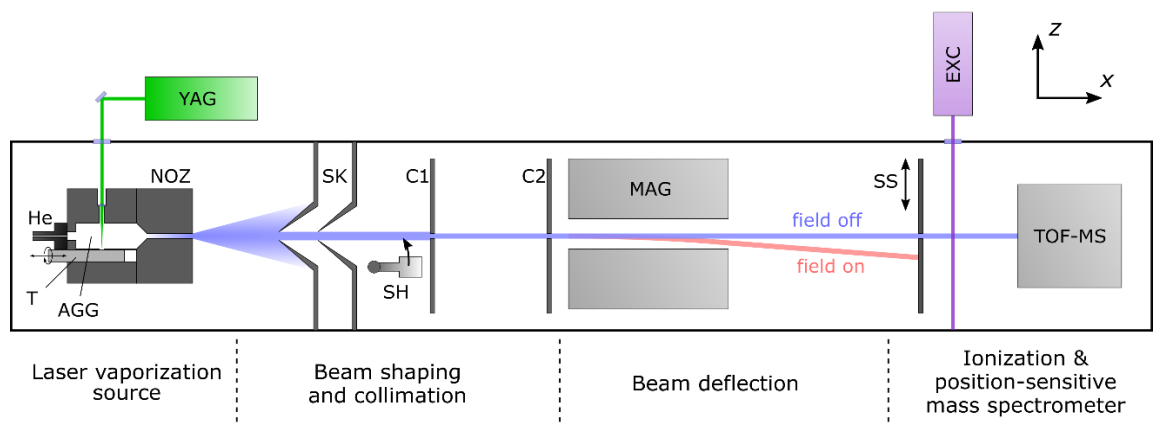

**Figure 3.** Outline of the magnet calibration experiment. NOZ: nozzle, He: helium pulsed valve, YAG: frequency-doubled Nd:YAG laser, T: target rod, AGG: aggregation chamber, SK: double skimmer, SH: shutter, C1 and C2: collimators, MAG: deflection magnet, SS: scanning slit aperture, EXC: $F_2$ excimer laser, TOF-MS: time-of-flight mass spectrometer.

The deflection of an atomic beam is given by[27]

$$d_z(M_J) = -\frac{(\frac{1}{2}l_1^2 + l_1 l_2)}{mv^2} g_J \mu_B M_J \frac{dB}{dz}. \qquad (1)$$

Here $m$ is the atomic mass, $M_J$ is the total magnetic angular momentum quantum number, $g_J$ is the Landé factor, $\mu_B$ is the Bohr magneton, $v$ is the atomic beam velocity, $l_1$ is the full length of the magnet, and $l_2$ is the field-free flight distance between the magnetic field and the beam detector. For Bi atoms $J = 3/2$ and $g_J = 1.6433$.[28]

The atomic beam was generated by laser evaporation[29] of a Bi rod. Its average velocity, measured with a mechanical shutter system,[30] was 1415 m/s.[31] At the end of the free flight path, the beam deflection profile was mapped out by a 0.4 mm-wide movable aperture. The atoms which passed through the aperture were ionized by 157 nm light from an excimer laser and then recorded by a time-of-flight mass spectrometer with a multi-channel plate detector.

Fig. 4 shows a comparison of Bi beam profiles produced by the two magnets. The zero-field profile and the dashed profile were recorded with the electromagnet. It has a two-wire geometry with $B_z$=1.3 T at 14 A current, and dimensions of $l_1$ = 14 cm and $l_2$ = 86 cm. Under the influence of the magnetic field the beam separates into four components ($M_J = \pm 1/2, \pm 3/2$), and from the parameters given above and a measured splitting of 0.87 mm, a field gradient of 305±18 T/m) is deduced.

Next, the electromagnet is replaced by the permanent magnet at nearly the same location ($l_1$ = 15 cm, $l_2$ = 84.3 cm). The separation of the atomic beam into four components is again clearly visible and is even larger, 1.01 mm, see the solid profile in Fig. 4. From the ratio of the beam splittings in the two cases we determine the permanent magnet deflector's field gradient to be 334± 20 T/m.

Like the magnetic field itself, the modeled value (508 T/m) is 35% higher than the measured value, as mentioned above. This suggests that by using custom-treated rather than off-the-shelf permanent magnets it could be realistic to produce even larger fields and gradients. Yet already at its current strength the deflector compares very favorably with the present and other electromagnets used for cluster beam experiments (see, e.g., the list in Ref. 32).

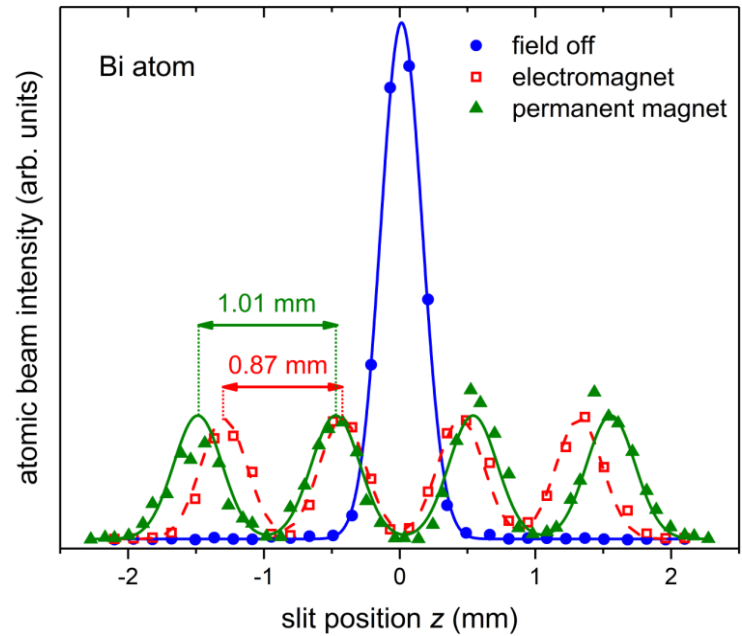

**Figure 4.** Magnetic deflection experiments on an atomic Bi beam using electromagnet (dashed red line) and permanent magnet (solid green line) deflectors. The beamlet data points for each measurement are fitted with four identical equidistant Gaussian functions. With the apparatus dimensions given in the text and Eq. (1), field gradients of 305 T/m and 334 T/m are deduced for the electro- and permanent-magnet units, respectively.

One may notice that the individual beamlets produced by the permanent magnet deflector are broadened compared to the other measurement, and display a small right-to-left loss in intensity. This of course is due not to the difference in the energizing magnets but to the different choices for their pole shoe geometries: two-wire vs. quadrupole-sector. The latter possesses a higher gradient orthogonal to the deflection direction, resulting in somewhat stronger and more inhomogeneous defocusing of the beam.[18,19] However, many experiments are concerned not as much with the homogeneity of focusing as with the strength and uniformity of the sideways ($z$-direction) deflecting force across the beam profile. For example, as mentioned in the Introduction, the pole shape chosen for this particular unit is based on its intended use for magnetic deflection experiments on ultra-cold paramagnetic atoms and molecules embedded in superfluid helium nanodroplets. In such experiments the variation of droplet masses[12,33] is of greater consequence than the slight variation in the defocusing component of the field.

## V. SUMMARY

In summary, we have built and tested a deflector for Stern-Gerlach-type experiments which is energized by using permanent magnets. The construction is rather straightforward, cost-efficient, compact, and transportable, while producing a magnetic field of 1.1 T and a field gradient of over 330 T/m in the molecular beam region, which is fully comparable with electromagnetic deflectors employed in research on atomic and molecular clusters. The described configuration has a quadrupole-type magnetic field, but the design is fully general and can be adapted to other pole shoe geometries. Another useful feature of the deflector concept demonstrated here is its scalability. For example, it is straightforward to increase its length: this requires only a longer magnet stack rather than more massive windings, power supply, and cooling system as in the case of an electromagnet. Conversely, a smaller-scale version can be constructed using smaller and cheaper magnets.[34]

**ACKNOWLEDGMENTS**

We would like to thank R. Agustsson of RadiaBeam Technologies, P. Musumeci of UCLA, and A. Robinson of SM Magnetics for useful advice, and S. Wieman and A. Flores of the USC Machine Shop for exceptional technical help and skillful machining. This work was supported by the U. S. National Science Foundation through Grant No. CHE-1664601, by the Deutsche Forschungsgemeinschaft through Grant No. SCHA885/17-1, and by the USC Undergraduate Research Associates Program.

**Data availability**: Data are available in the article and upon request from the authors.